\documentclass[
 reprint,
 amsmath,amssymb,
 aps,
]{revtex4-2}

\usepackage{graphicx}
\usepackage{dcolumn}
\usepackage{bm}
\usepackage{blindtext}
\usepackage{hyperref}
\usepackage{graphicx}
\usepackage{xcolor}
\usepackage{dsfont}

\begin{document}

\preprint{APS/123-QED}

\title{Learning locally dominant force balances in active particle systems}

\author{Dominik Sturm$^{1,2,3}$, Suryanarayana Maddu$^{4,5}$, Ivo F.~Sbalzarini$^{1,6,7,8,9}$} \email{sbalzarini@mpi-cbg.de}
\affiliation{$^{1}$ Faculty of Computer Science, Technische Universit\"{a}t Dresden, Dresden, Germany}
\affiliation{$^{2}$ Center for Advanced Systems Understanding (CASUS), G\"orlitz, Germany}
\affiliation{$^{3}$ Helmholtz-Zentrum Dresden-Rossendorf, Dresden, Germany}
\affiliation{$^{4}$ Center for Computational Biology, Flatiron Institute, New York, USA}
\affiliation{$^{5}$ NSF-Simons Center for Mathematical \& Statistical Analysis of Biology, Harvard University, Boston, USA}
\affiliation{$^{6}$ Max Planck Institute of Molecular Cell Biology and Genetics, Dresden, Germany}
\affiliation{$^{7}$ Center for Systems Biology Dresden, Dresden, Germany} 
\affiliation{$^{8}$ Center for Scalable Data Analytics and Artificial Intelligence ScaDS.AI, Dresden/Leipzig, Germany}
\affiliation{$^{9}$ Cluster of Excellence Physics of Life, Technische Universit\"{a}t Dresden, Dresden, Germany}

\date{\today}

\begin{abstract}
We use a combination of unsupervised clustering and sparsity-promoting inference algorithms to learn locally dominant force balances that explain macroscopic pattern formation in self-organized active particle systems. The self-organized emergence of macroscopic patterns from microscopic interactions between self-propelled particles can be widely observed nature. Although hydrodynamic theories help us better understand the physical basis of this phenomenon, identifying a sufficient set of local interactions that shape, regulate, and sustain self-organized structures in active particle systems remains challenging. We investigate a classic hydrodynamic model of self-propelled particles that produces a wide variety of patterns, like asters and moving density bands. Our data-driven analysis shows that propagating bands are formed by local alignment interactions driven by density gradients, while steady-state asters are shaped by a mechanism of splay-induced negative compressibility arising from strong particle interactions. Our method also reveals analogous physical principles of pattern formation in a system where the speed of the particle is influenced by local density. This demonstrates the ability of our method to reveal physical commonalities across models. The physical mechanisms inferred from the data are in excellent agreement with analytical scaling arguments and experimental observations.
\end{abstract}

\maketitle

\section{Introduction}

Systems of self-propelled particles can exhibit self-organized collective behavior that leads to the formation of complex spatio-temporal patterns. This phenomenon is ubiquitous in nature and can be observed across all scales, ranging from the active self-assembly of the mitotic spindle by microtubule and motor proteins \cite{shelley_dynamics_2016} and the formation of rich vortex structures in dense bacterial suspensions \cite{wensink_meso-scale_2012} to the directed and collective motion of cells in tissues \cite{munster_attachment_2019} and organoids \cite{tan_emergent_2022} up to fish shoals and flocks of birds \cite{toner_flocks_1998}. Recent interest has also gone towards the self-assembly of active and adaptive materials \cite{lam_cytoskeletal_2016, needleman_active_2017} and the use of microswimmers in biomedical applications \cite{bechinger_active_2016, bunea_recent_2020}. Despite the prevalence of self-organizing systems composed of self-propelled particles, it remains difficult to identify a sufficient set of mechanisms that shape and regulate pattern formation in active particle systems \cite{shelley_dynamics_2016}.

First insights into this type of self-propelled particle systems were obtained by Vicsek \textit{et al.}~\cite{vicsek_novel_1995}. They studied a minimal microscopic model of self-propelled particles at constant speed with local alignment interactions. At high particle density and low noise, they reported a distinct phase transition from a unordered state to a flocking state with global order, where the dynamics of the system are primarily determined by polar alignment interactions. Later, derivations of hydrodynamic theories based on symmetry arguments, for example by Toner and Tu \cite{toner_long-range_1995, toner_flocks_1998}, gave a mean-field perspective on the physical mechanisms. Since then, both microscopic and macroscopic systems have been shown to be able to form a wide range of spatio-temporal patterns that may also be observed experimentally  \cite{gopinath_dynamical_2012}, including asters \cite{ndlec_self-organization_1997,surrey_physical_2001} and moving density bands \cite{schaller_polar_2010}. In parallel, theories derived by analytical coarse-graining of the microscopic model \cite{bertin_hydrodynamic_2009} resulted in similar mean-field descriptions, where the parameters of the coarse-grained model could be directly linked to the strength of the microscopic interactions \cite{saintillan2013active}.

In the recent years, machine learning has found wide applicability in physical and life sciences \cite{brunton_discovering_2016,rudy_data-driven_2017,brunton_machine_2020}. Data-driven techniques have been successfully applied to infer ordinary and partial differential equation (PDE) models from observation data \cite{mangan_inferring_2016,maddu_stability_2022} and for algorithmic coarse-graining of microscopic systems of self-propelled particles \cite{supekar_learning_2023, maddu_learning_2022, joshi_data-driven_2022}. In a different line of work, neural networks have been used to infer the activity and temporal evolution of a system for a given orientation field in nematohydrodynamics \cite{colen_machine_2021}. Similarly, physics-informed neural networks, which use an assumed physical model in the form of an ODE/PDE to structurally regularize a deep neural network \cite{raissi_hidden_2020}, have been used to predict model parameters and effective pressures for turbulent flows in dense bacterial suspensions \cite{maddu_inverse_2022}. Finally, recent work has shown that reduced-order models, like proper orthogonal decomposition (POD), can be used to extract coherent flow structures of velocity fields in bacterial turbulence under various conditions \cite{henshaw_dynamic_2023}.

Although data-driven methods have proven successful at inferring macroscopic models or  extracting coherent flow structures, they generally capture the average global dynamics of the system. This is because they infer a global model that sufficiently describes the data everywhere. However, they do not capture local differences in the microscopic interactions that can explain the emergence of macroscopic heterogeneous patterns. What one would need for that is a method that infers a domain decomposition of the system into (a minimal set of) local neighborhoods of similar dynamics and identifies the relative importance of the terms of a global model in each local neighborhood. These terms constitute the locally dominant force balances. They are sufficient to explain the dynamics within a local neighborhood and provide insight into the local contributions of the different terms that drive the dynamics in the region. If in addition the global model is derived by analytical coarse-graining of a microscopic model, it allows one to learn the microscopic origin of the observed dominant macroscopic principles.

We use an algorithm that builds on the work by \cite{callaham_learning_2021}, who used a combination of unsupervised clustering and a sparse principle components analysis (SPCA) to identify the dominant components of a global model that drive the spatiotemporal dynamics in a predefined local region. This is different from typical sparse regression approaches used for model identification, as the goal here is to identify the local relative importance of terms \cite{constantine_active_2014} in a predefined region equipped with the full knowledge of the global model parameters. In this study, we present an approach that extends the algorithm \cite{callaham_learning_2021} by an improved model-selection process based on computing more than just the leading principle component to obtain clear Pareto fronts.

We then apply the data-driven method to the hydrodynamic mean-field description of a self-propelled particle system with alignment interactions. The numerical experiments demonstrate that the data-driven approach is capable of identifying  mechanisms of pattern formation that agree with analytical predictions obtained from asymptotic and linear stability analysis. By then considering a slightly different model with density-dependent motility, we report the identification of similar mechanisms of pattern formation. This suggests that the present data-driven approach is robust enough to identify common physical principles across active particle systems, even when the underlying microscopic rules are different. Therefore, we believe that the presented approach can be used in a wide range of active matter systems to probe the microscopic physical mechanisms driving the emergence of self-organized structures.

\section{Methods}\label{sec:method}

We review the hydrodynamic mean-field model of an active particle system, followed by a brief description of the data-driven approach used for learning locally dominant force balances.

\subsection{Hydrodynamic Model of Self-Propelled Particles}\label{sec:method:model}

We consider a classic system of self-propelled point particles with polar alignment interactions that move at a speed $w_0$ in the overdamped viscous limit~\cite{vicsek_novel_1995}. In the absence of ``leader particles'' and external forces, this model is capable of a phase transition to an ordered flocking state with increasing particle density, highlighting the self-organized nature of the system \cite{gregoire_onset_2004}.
The hydrodynamic mean-field description of this system has been derived both from first principles~\cite{toner_long-range_1995, toner_flocks_1998} and by analytic coarse-graining~\cite{bertin_hydrodynamic_2009, farrell_pattern_2012, marchetti_hydrodynamics_2013}. The corresponding PDEs describe the space-time evolution of a mass-conserved particle number density $\rho$ and the polarity density $\bm{W}=\rho\bm{P}$, where the polarity $\bm{P}$ is an order parameter of the system describing the average local orientation of particles. The polarity density $\bm{W}$ hence both describes the local order in the system as well as the velocity field by which the number density $\rho$ is  advected. This dual role is crucial in determining the large-scale behavior of the system \cite{marchetti_hydrodynamics_2013}. Here, we consider the model following Gopinath \textit{et al.} \cite{gopinath_dynamical_2012} that is given by:
\begin{align}\label{eq:model:continuoum_model:density}
\partial_t\rho =& -\nabla\cdot(w_0\bm{W} - D\nabla\rho)\\\nonumber\\\label{eq:model:continuoum_model:polarity}
\partial_t\bm{W} + \lambda_1(\bm{W}\cdot\nabla)\bm{W} =& -[a_2(\rho)+ a_4(\rho)|\bm{W}|^2]\bm{W}\\
&+ D_W\Delta\bm{W} - w_0\nabla\rho \nonumber \\
&+ \frac{\lambda_3}{2}\nabla|\bm{W}|^2 + \lambda_2\bm{W}(\nabla\cdot\bm{W}).\nonumber
\end{align}
Equation~\eqref{eq:model:continuoum_model:density} is an advection-diffusion equation transporting $\rho$ with the self-propulsion speed $w_0$ in the direction of the local particle orientation $\bm{P}$. The scalar parameter $D$ is a diffusion constant to account for the thermal noise in the microscopic model. 

Equation \ref{eq:model:continuoum_model:polarity} is the transport equation for the polarity density, encapsulating the mean-field effect of the alignment interactions between the particles. It is a generalized Navier-Stokes-type equation, where the convective nonlinearity associated to $\lambda_1$ accounts for the self-advection of the particles and is determined by the microscopic properties of the model \cite{marchetti_hydrodynamics_2013}. In fact, this model is the first-order truncation of the classic Toner-Tu model keeping only one term on the order of diffusion. The first term on the right-hand side of Eq.~\eqref{eq:model:continuoum_model:polarity} models spontaneous polarization of moving particles. The scalar $D_W$ is a diffusion constant describing the alignment interactions between particles as relaxation of the splay and bend moduli of the polarity density field. The third and fourth terms on the right-hand side can be interpreted as the effective hydrodynamic pressure, where $w_0\nabla\rho$ corresponds to the ideal-gas part and the term associated to $\lambda_3$ accounts for the pressure induced by the splay of the polarity field, which again depends on the microscopic interactions between he particles~\cite{gopinath_dynamical_2012}. The last term associated with $\lambda_2$ accounts for the nonlinear feedback between the polarity density $\bm{W}$ and the compressibility of the flow \cite{bertin_hydrodynamic_2009}. Following Gopinath {\it et al.}~\cite{gopinath_dynamical_2012}, we consider a parameterization in which the self-propulsion speed $w_0$ defines the scaling of both the particle movement and the ideal-gas part.  

Although there exist more elaborate active particle models \cite{bechinger_active_2016}, the above minimal model is able to capture several universal characteristics of active flows, such as the onset of global order, large density fluctuations, and spatio-temporal pattern formation \cite{farrell_pattern_2012}.
Indeed, depending on the choice of the parameters $w_0$
and $\lambda_i$, Eqs.~\eqref{eq:model:continuoum_model:density} and \eqref{eq:model:continuoum_model:polarity} exhibit several instabilities of the polarized homogeneous steady state. 
The polarity playing the dual role of both a local order parameter and the density-advection velocity provides the necessary nonlinear feedback for these instabilities, as the density controls the orientational order and is itself advected by the order parameter \cite{gopinath_dynamical_2012}. This leads to the emergence of macroscopic structures, such as moving density bands or radially symmetric asters, which have also been observed in living systems \cite{schaller_polar_2010, ndlec_self-organization_1997, surrey_physical_2001}. Many works have identified the physical mechanisms of pattern formation in these hydrodynamic equations \cite{mishra_fluctuations_2010, gopinath_dynamical_2012, bertin_hydrodynamic_2009,sankararaman_self-organized_2004, gowrishankar_nonequilibrium_2016}.
Yet, it remains difficult to determine the microscopic origins of those mechanisms~\cite{shelley_dynamics_2016}, as even if the underlying model is known, it is hard to identify a locally sufficient set of physical processes that can explain the formation of macroscopic spatio-temporal patterns.

\subsection{Learning Locally Dominant Force Balances}\label{sec:method:inference}

We learn the locally dominant force balances for the dynamical system in Eqs.~\eqref{eq:model:continuoum_model:density} and \eqref{eq:model:continuoum_model:polarity} following the approach introduced by Callaham {\it et al.}~\cite{callaham_learning_2021}. It is based on assuming that the space-time dynamics of an intensive scalar field $u(\bm{x},t)$ is given by
\begin{equation}\label{eq:appendix:methods:dynamical_system}
\partial_tu(\bm{x}_i) + \mathcal{N}(u(\bm{x}_i);\bm{\xi}) = 0\, ,
\end{equation}
at discrete space-time points $\bm{x}_i=(\bm{x}_i, t_i)$, $i=0,\dots,N$, sampled from a domain $\bm{x}_i\in\Omega\subseteq\mathbb{R}^n$, $t_i\in[0,T_{\max}]$. The nonlinear right-hand side $\mathcal{N}$ contains the $F$ differential operators with known coefficients $\bm{\xi}=(\lambda_1, w_0, \dots)$ of the given PDE model, thus
\begin{equation}
    \mathcal{N}(u(\bm{x}_i);\bm{\xi}) = \sum\limits_{j=1}^{F} \bm{\xi}_jf_j(u(\bm{x}_i))\, .
\end{equation}
For Eq.~\eqref{eq:model:continuoum_model:polarity}, $f_1(\bm{x}_i)=(\bm{W}(\bm{x}_i)\cdot\nabla)\bm{W}(\bm{x}_i)$, and so on, such that $F=7$. 

Given the $f_j$ alongside $\bm{\xi}_j$ and corresponding measurement or simulation data, we construct the feature matrix $\Theta$ as
\begin{equation}\label{eq:appendix:methods:feature_matrix}
    \Theta = \begin{bmatrix}
        \partial_t \bm{W}_x(\bm{x}_1) & \bm{\xi}_1f_1(\bm{x}_1) & \cdots & \bm{\xi}_{F}f_{F}(\bm{x}_1) \\
        \vdots & \vdots & \ddots & \vdots \\
        \partial_t \bm{W}_x(\bm{x}_N) & \bm{\xi}_1f_1(\bm{x}_N) & \cdots & \bm{\xi}_{F}f_{F}(\bm{x}_N)
    \end{bmatrix}\in\mathbb{R}^{N\times (F+1)}.
\end{equation}
The columns of $\Theta$ span the so-called equation space \cite{callaham_learning_2021}. By construction, $\bm{1}^\top\bm{\theta}_i=0$ for each equation-space sample (i.e., row of $\Theta$) $\bm{\theta}_i=\Theta_{i,:}$. As $\pmb{W}$ has two spatial components, we stack the $x$- and $y$-components of the polarity vertically in order to infer locally dominant forces for both $x$ and $y$. Therefore, $\Theta=[\Theta_x, \Theta_y]^\top$ where the subscripts denote the corresponding spatial components of the vector quantities.

Inferring locally dominant force balances then corresponds to restricting the dynamics to some subspace spanned by the axes of highest variance in the data \cite{callaham_learning_2021}. In this interpretation we neglect terms of small local variance, as their contribution to the dynamics are minor. Thus, the approach amounts to a localized active subspace method in equation space \cite{constantine_active_2014}.

To identify regions of similar covariance, we feed columns of the matrix $\Theta$ as input features to the Gaussian Mixture Model (GMM) to output $K$ clusters $c_k$, $k=1,\ldots ,K$. These clusters thus correspond to a set of points in space and time 
that exhibit similar dynamics. These regions could alternatively be identified using other clustering approaches, but GMMs provide information about the local contributions of the functions $f_i$ in each cluster $c_k$ via the fitted covariance matrices. This is valuable information, since a higher covariance along the equation-space axis $i$ in cluster $k$ indicates that the model term $f_i$ contributes more to the local dynamics in cluster $c_k$. This allows identifying the set of {\em dominant} terms that constitutes the locally dominant force balance at all space-time points $\bm{x}_i \in c_k$.

Deciding whether a term $f_i$ is dominant in a given cluster $c_k$ is done by thresholding the covariance.
Finding the optimal threshold is a difficult problem, especially since the covariance matrices of all GMM components are dense. Callaham {\it et al.}~\cite{callaham_learning_2021} therefore proposed to only look at a sparse approximation by the principle component that explains most of the observed variance over $\Theta_k$ using sparse principle components analysis (SPCA). This provides a sparse and interpretable representation of the axis of maximum variance to which the equation space can be restricted. Here, we take into account more than just one principle dimension by solving the SPCA problem as a piecewise-convex dictionary learning problem \cite{mairal_online_2009}:
\begin{align}\label{eq:appendix:methods:objective}
    \hat{C}_k(\alpha), \hat{D}_k(\alpha) = &\arg\min_{C,D}\frac{1}{2}\big\Vert\widetilde\Theta^\top _k - CD\big\Vert^2_F + \alpha\big\Vert C\big\Vert_1\\
    &\mathrm{s.t.}\text{ } \big\Vert D_{i,:}\big\Vert_2 \leq 1,\label{eq:appendix:methods:objective:constraint}
\end{align}
where $\widetilde{\Theta}_k=\Theta_k - \overline{\Theta}_k$, $\widetilde{\Theta}_k^\top=V\Sigma U^\top$ is the singular value decomposition (SVD), and $\overline{\Theta}_k$ the column-wise mean of $\Theta_k$. Moreover, $C=V_{:r}\in\mathbb{R}^{(F+1)\times r}$ and $D=(\Sigma U^\top)_{:r}\in\mathbb{R}^{r\times N_k}$ are the rank-$r$ truncated approximations of the SVD of $\widetilde{\Theta}^\top$~\cite{brunton_data-driven_2019}.
Throughout this paper we chose $r=6$ as determined by the ``elbow'' method on the reconstruction error of Eq.~\eqref{eq:appendix:methods:objective} which is a common approach for determining a point of diminishing returns~\cite{satopaa_finding_2011}. The constraint in Eq. \eqref{eq:appendix:methods:objective:constraint} is enforced to ensure that $C$ does not become arbitrarily small, but also means that the resulting matrices are no longer guaranteed to be orthogonal \cite{mairal_online_2009}. The term associated to $\alpha$ enforces that the reconstruction of $\hat{C}_k(\alpha)$ is sparse w.r.t.~the matrix $L_1$-norm, i.e. over the elements in $C$. We use the implementation of the entire algorithm provided by $\textsc{sklearn}$~\cite{pedregosa_scikit_2011}.

The presented SPCA-based active subspace selection is in stark contrast to global hard thresholding of the operators when determining the set of active processes. As we are interested in identifying the {\em locally} dominant forces in the system, the scales of importance differ locally \cite{callaham_learning_2021}. Therefore, finding locally optimal thresholds would be a combinatorially hard problem, which would require exhaustive search without guarantee to reflect the dynamics in the data. In contrast, the present method automatically detects local regions of similar dynamics in a data-driven fashion, which reduces the thresholding problem to identifying a sparsity parameter $\alpha$ for the SPCA, which can be done in a principled fashion using concepts from Pareto-optimality \cite{mangan_inferring_2016}.

\subsection{Model Selection}\label{sec:method:selection}

Looking at the optimization problem in Eq.~\eqref{eq:appendix:methods:objective} makes clear that the solution will depend on the choice of the sparsity-promoting parameter $\alpha$. It is used to balance the accuracy of the model against its complexity, acting as an inductive bias that assumes the dynamics can be locally well approximated by the balance of few dominant terms. For some $\alpha_{\min}$ we obtain $\hat{C}_k(\alpha_{\min})\approx V_{:r}$, $0>r\leq F+1$ with a low residual, while a value above $\alpha_{\max}$ results in the sparsest (but trivial) model $\hat{C}_k(\alpha_{\max})=\bm{0}$ with a high reconstruction error. While we are able to determine $\alpha_{\max}$ analytically for classic LASSO problems with convex penalties \cite{maddu_stability_2022}, such estimates cannot be obtained for Eq.~\eqref{eq:appendix:methods:objective}, as it depends on a dictionary $D$. Thus, determining $\alpha$ requires repeated evaluation of the objective over a wide range of $\alpha$ values, resulting in a so-called $\alpha$-\textit{path} that can be used for model selection. 

We construct the regularization path over a fixed interval $[\alpha_{\min}, \alpha_{\max}]$ containing 400 log-equidistant points to achieve sufficient resolution. The interval bounds (here: $\alpha_{\min}=10^{-4}$, $\alpha_{\max}=10^2$) are empirically determined to obtain a full Pareto front, going from a full to a nearly empty set of model terms (see Appendix~\ref{sec:exp_details}). They are kept the same for all experiments of a given model, as the bounds achieve the desired behavior over all considered model coefficient values. For each cluster $c_k$, we evaluate the reconstruction error, denoted by $\epsilon_k(\alpha)$, along the $\alpha$-path as the residual of Eq.~\eqref{eq:appendix:methods:objective}. The Pareto front for model selection can be visualized by plotting the reconstruction error $\epsilon_k(\alpha)$ versus the number of non-zero elements (active terms) identified.

The above model selection procedure determines a potentially different $\alpha$ for each cluster $c_k$, since the required penalization strength depends on the cluster-specific data. This is different from the global $\alpha$ used by Callaham {\it et al.}~\cite{callaham_learning_2021}, identified as the ``elbow'' over the squared $L_2$-norm of the ignored terms globally. We found the global criterion difficult in practice, as it was hard to identify a definitive selection point in many cases or might result in nearly empty supports. While the use of a global $\alpha$ might speed up the model selection procedure, it would also ignore the local relative importance of PDE terms, since the global $\alpha$ is dominated by the GMM components of largest magnitude. We also find that the use of the reconstruction error as selection criterion is justified, as it measures how well the structure of the data in equation space is approximated by the chosen principle directions $\hat{C}_k(\alpha)$. 

\section{Results}\label{sec:results}

We apply the presented method to infer locally dominant force balances underlying the formation of moving bands and asters obtained from numerical simulations of the hydrodynamic model in Eqs. \eqref{eq:model:continuoum_model:density} and \eqref{eq:model:continuoum_model:polarity}. As discussed in the previous section, we first decompose the spatiotemporal domain into distinct non-overlapping clusters using GMM. For each region marked by the cluster index, we employ SPCA to infer locally dominant components of the global model driving spatiotemporal organization in the localized region. We apply our inference strategy to the hydrodynamic models with and without density-dependent motility to assess the role of varying particle speed in shaping the emergent patterns.

\subsection{Moving Density Bands}\label{sec:results:parallel}

\begin{figure*}
    \centering
    \includegraphics[width=\linewidth,trim={0 0 4.5cm 0},clip]{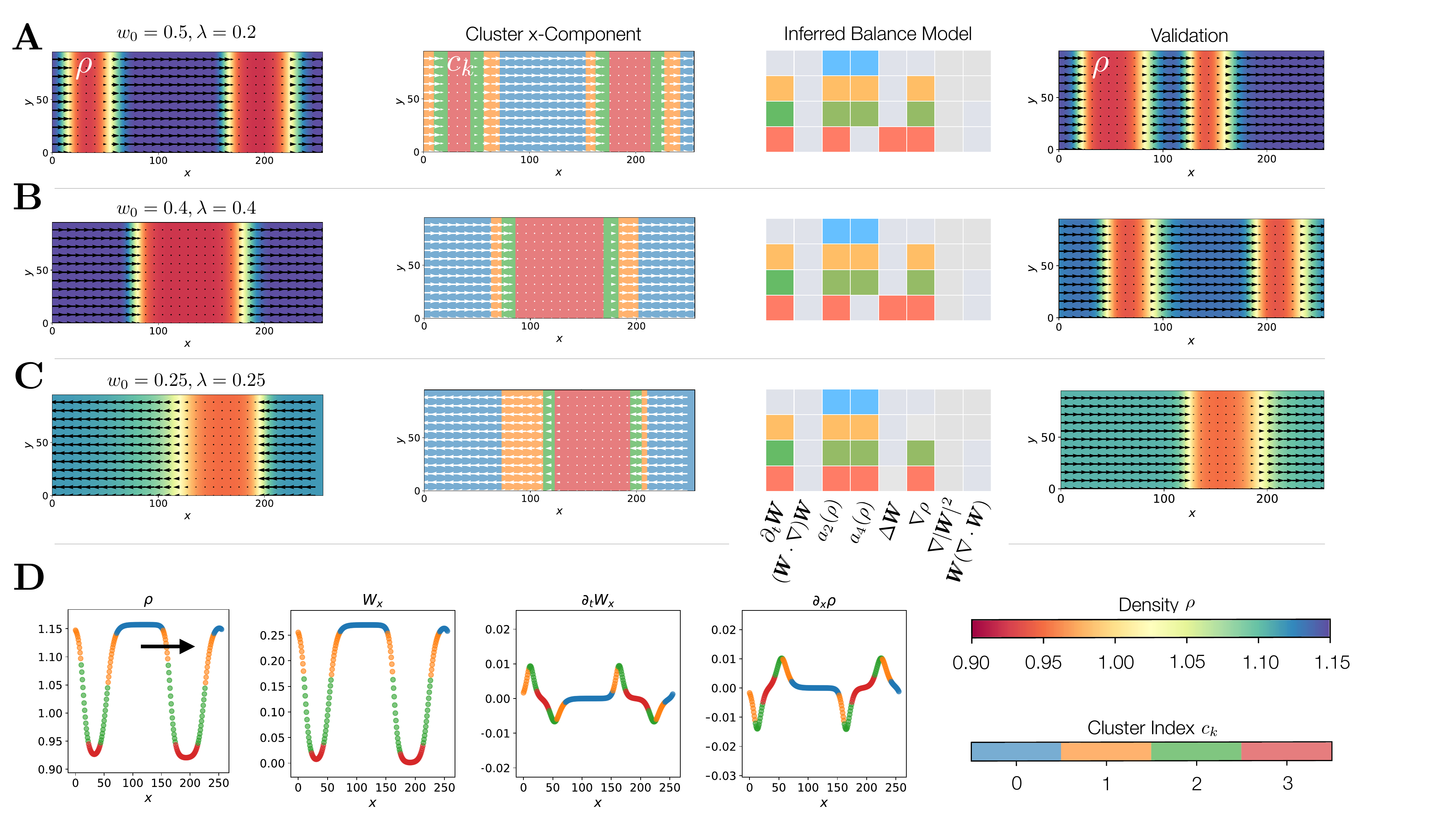}
    \caption{Clusters and the corresponding local dominant models for three different parameterizations of Eqs.~\eqref{eq:model:continuoum_model:density} and \eqref{eq:model:continuoum_model:polarity} in the convection-dominated regime forming moving density bands. {\bf{(A--C)}} Each row corresponds to a parameter set (A: $w_0=0.5$, $\lambda=0.2$; B: $w_0=0.4$, $\lambda=0.4$; C: $w_0=0.25$, $\lambda=0.25$). Reading from left to right: (1) visualization of the data in the two-dimensional simulation domain $[0,256]\times [0,96]$ at the last data point in time (color: number density $\rho$, arrows: polarity density $\bm{W}$); (2) identified clusters of similar dynamics for the $x$-component of $\bm{W}$ (color: cluster index $c_k$); 
    (3) inferred dominant force-balance models in each cluster (Filled squares highlight the active components and colors correspond to different clusters); (4) numerical validation of the minimal pattern-forming model in Eq.~\eqref{eq:results:parallel:minimal_dynamic} starting from the same initial condition. 
    {\bf{(D)}} Line profiles of the data ($\rho$, $\bm{W}$, and temporal derivative of $\bm{W}$ and $\partial_x\rho$) along $x$ (for $y=20$), colored by cluster membership, for $w_0=0.5$, $\lambda=0.2$. The arrow indicates the direction of motion of the density bands.}
    \label{fig:results:parallel:overview}
\end{figure*}

We first consider the self-organized dynamics of moving density bands that manifest as wave-like solutions of the hydrodynamic PDEs. These types of structures can also be observed in \textit{in-vitro} reconstitution of self-propelled molecules, such as in dense actin motility assays, where moving band-like structures emerge and persist over several minutes \cite{schaller_polar_2010}. 

We numerically solve for Eqs.~\eqref{eq:model:continuoum_model:density} and \eqref{eq:model:continuoum_model:polarity} (see Appendix \ref{sec:simulation}) with $\lambda_1=\lambda_2=\lambda_3=\lambda$, for three different parameterizations of the model PDEs (Fig.~\ref{fig:results:parallel:overview}A-C). In the numerical solution an initially homogeneous state starts forming high density spots that move in the direction of mean orientational order. Through diffusion the regions expand normal to the direction of motion until they span the length of the simulation domain \cite{gopinath_dynamical_2012}. This results in the formation of high-density bands that move through an disordered background at a speed that linearly scales with with the microscopic self-propulsion speed $w_0$ \cite{mishra_fluctuations_2010}. Different realizations of these non-equilibrium steady-state structures can be seen in Fig.~\ref{fig:results:parallel:overview}A--C for the same initial condition, but shown for three different parameters of the model. The ordered high-density regions are enclosed by two sharp wave fronts in both density and polarity, as seen in Fig.~\ref{fig:results:parallel:overview}D for a density band moving in the direction denoted by the arrow. 

In the data of the numerical solutions, we identify the time interval in which the high-density structure is formed and advected in a stable manner over an extended time horizon (around 50\,000 simulation time steps). Multiple time points are collected as described in Appendix \ref{sec:simulation} and the data used to compute the feature matrix $\Theta$. The spatio-temporal samples $\bm{\theta}_i=\Theta_{i,:}$ are decomposed into $K_{\mathrm{band}}=4$ clusters by the GMM. The identified clusters of similar covariance, and hence similar local dynamics, are shown in the second column of Fig.~\ref{fig:results:parallel:overview}A--C, where the cluster index (represented by color) increases from the high-density regions to the low-density regions. We find that the clusters decompose the spatio-temporal domain into well-defined  symmetric regions, where cluster 0 contains the ordered high-density phase, 1 and 2 the sharp wave fronts at the interfaces between the two phases, and 3 the unordered background. As the solution is invariant in the $y$-direction, i.e.~$\partial_t \bm{W}_y = 0$, the $y$-components of $\bm{W}$ are always assigned to the unordered background cluster (not shown). Moreover, we find that clusters 1 and 2 divide the wave fronts roughly at the inflection points of the polarity density $\bm{W}$, as seen in Fig.~\ref{fig:results:parallel:overview}D, thus capturing the local gradients of the spatial and temporal dynamics.

Performing SPCA in each of the identified clusters, we find a sparse approximation on the local directions of maximal variance according the Pareto fronts shown in Fig. \ref{fig:results:parallel:pareto}. The resulting locally dominant force balances are shown in the third column of Fig.~\ref{fig:results:parallel:overview}A--C. For all three parameterizations of the PDE, comparable local mechanisms are identified that depend on the microscopic interactions through $a_2(\rho)$ and $a_4(\rho)$
and on the density gradient $\nabla\rho$. This shows the robustness of the method against perturbations in the system parameters. In fact, the dynamics in the high-density regions (blue cluster 0 in Fig.~\ref{fig:results:parallel:overview}A--C) can be described by a steady-state equation depending on the spontaneous polarization terms as
\begin{equation}\label{eq:results:parallel:minimal_steady}
    0 = - [a_2(\rho) + a_4(\rho)|\bm{W}|^2]\bm{W}.
\end{equation}
This local model in cluster 0 is independent of the temporal evolution  $\partial_t\bm{W}$, since the change in polarity density is concentrated around the wave fronts (see Fig.~\ref{fig:results:parallel:overview}D). This observation is consistent with previous work \cite{gopinath_dynamical_2012}, since the high-density region can be interpreted as the homogeneous steady-state solution of Eqs.~\eqref{eq:model:continuoum_model:density} and \eqref{eq:model:continuoum_model:polarity}. For a spatially homogeneous polarity $\bm{W}$ and constant density $\rho_0$, as seen in Fig.~\ref{fig:results:parallel:overview}D, the system possesses two stable steady states: $\bm{P}=0$ if $\rho_0\leq\rho_c$ and $\bm{P}=\sqrt{-a_2(\rho_0) /a_4(\rho_0)}\bm{e}_\theta$ for $\rho_0>\rho_c$, where $\bm{e}_\theta$ is the direction of broken symmetry. At a microscopic scale, the polarization terms arise from the cooperative effect of many filaments interacting through alignment interactions, where at the critical density the transition to an ordered state occurs and all particles align \cite{bertin_hydrodynamic_2009}. As such, the homogeneously polarized case directly reflects the local force-balance model, where we observe the onset of collective motion through strong particle interactions.

The high-density regions are separated from the unordered (unpolarized) background by steep wave fronts, whose local dynamics are assigned into clusters 1 and 2 (orange and green in Fig.~\ref{fig:results:parallel:overview}). In these regions, the locally dominant force balance is given by the dynamic model
\begin{equation}\label{eq:results:parallel:minimal_dynamic}
    \partial_t \bm{W} = -[a_2(\rho)+ a_4(\rho)|\bm{W}|^2]\bm{W} - w_0\nabla\rho \, .
\end{equation}
The inferred model captures the order--disorder transition of the system through alignment interactions. This is reflected in the appearance of the spontaneous polarization terms $a_2(\rho)$ and $a_4(\rho)$, which lead to the emergence of the ordered phase above the characteristic density $\rho_c$~\cite{gopinath_dynamical_2012}. 
The importance of polar alignment interactions for the formation of the density bands has previously been identified \textit{in-silico} through particle simulations and \textit{in-vitro} experimental studies \cite{schaller_polar_2010}. The identified local model also captures correctly that the wave fronts are advected in the direction of mean orientational order due to the arising density gradient $\nabla\rho$. This coupling of $\bm{W}$ and $\rho$ gives rise to the convective nature of the ordered state \cite{gopinath_dynamical_2012, mishra_fluctuations_2010} and the density dependence is also captured in traveling-wave solutions derived from microscopic models \cite{bertin_hydrodynamic_2009}.
We also identify a similar driving mechanism in the disordered background (cluster 3), which includes an additional diffusive term.

Similarly to the data-driven model in Eq.~\eqref{eq:results:parallel:minimal_dynamic}, previous work has described the moving bands as a soliton solution, i.e.
\begin{equation}
     \partial_t\bm{W} = -[a_2(\rho)+ a_4(\rho)\frac{|\bm{W}|^2}{\rho^2}]\bm{W} - w_0\partial_x\rho \, ,
\end{equation}
where the nonlinear polarization terms provide the dispersion to generate the wave structure \cite{gopinath_dynamical_2012, bertin_hydrodynamic_2009}. The local dominant components identified by the data-driven (Eq.~\ref{eq:results:parallel:minimal_dynamic}) strategy agree with the aforementioned analytical model derived through asymptotic analysis \cite{bertin_hydrodynamic_2009}.

The local force balances inferred here can be linked to a convection-mediated density instability of the homogeneous steady state, which coincides with the onset of moving density bands \cite{gopinath_dynamical_2012}. This instability originates from the tendency of the system to build local order through $a_2$ and $a_4$, together with the convective coupling of $\rho$ and $\bm{W}$ through $w_0$. The resulting effective pressure difference leads to the advection of the ordered phase, as also identified here from data.

The model in Eq.~\eqref{eq:results:parallel:minimal_dynamic} contains all terms present in any of the clusters, except for the (weak) background diffusion term.
As such, it constitutes the minimal union model for this system. Numerically solving this minimal pattern-forming model confirms that it is sufficient to form moving density bands (see rightmost column of Fig.~\ref{fig:results:parallel:overview}A--C), despite being much simpler than the governing Eqs.~\eqref{eq:model:continuoum_model:density} and \eqref{eq:model:continuoum_model:polarity}. These are consistent with previous results showing that moving density bands form also in the absence of convective nonlinearities~\cite{gopinath_dynamical_2012}.

\subsection{Asters}\label{sec:results:aster}

\begin{figure*}
    \centering
    \includegraphics[width=\linewidth,trim={0 0 7cm 0},clip]{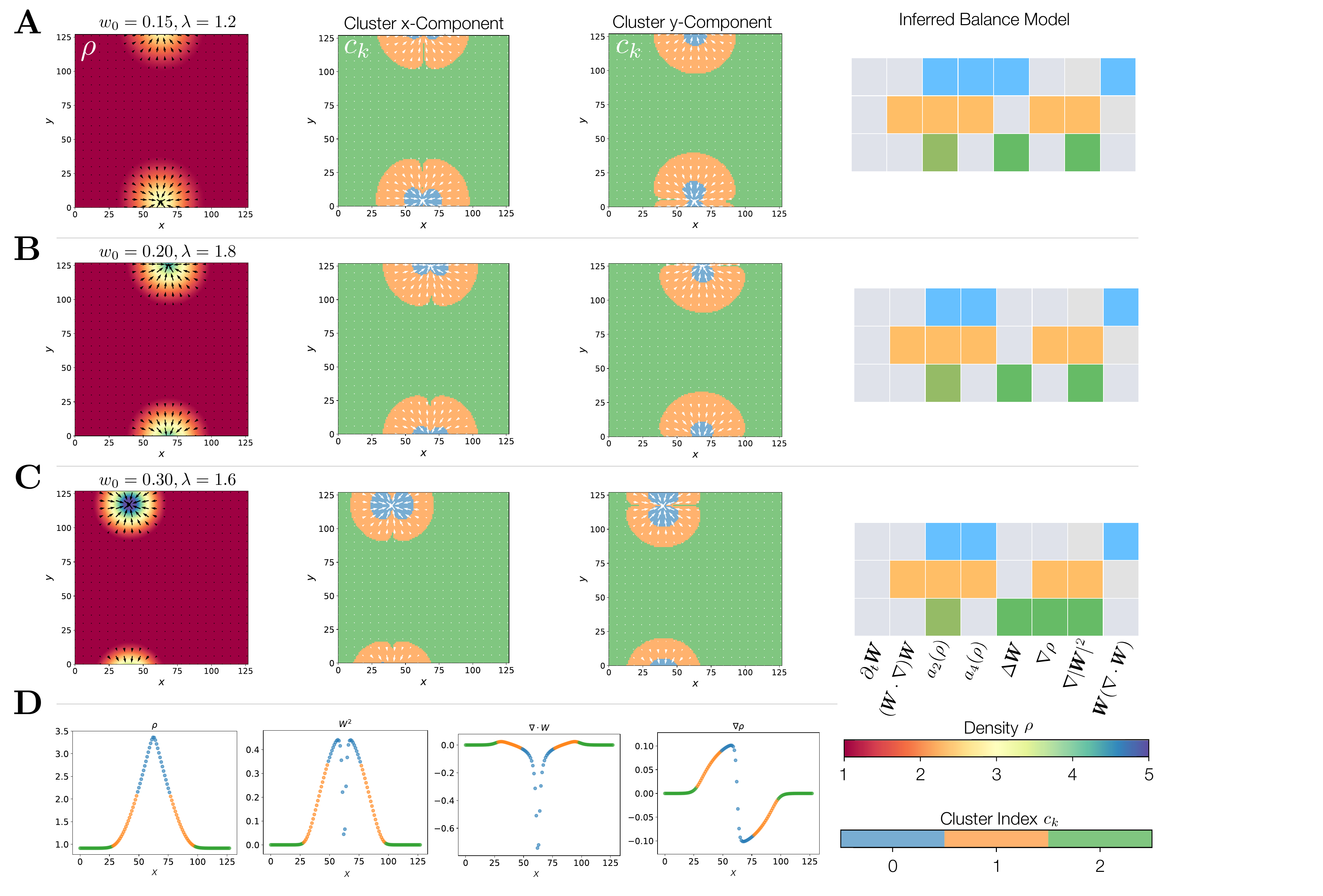}
    \caption{Clusters and the corresponding local dominant models for three different parameterizations of Eqs.~\eqref{eq:model:continuoum_model:density} and \eqref{eq:model:continuoum_model:polarity} forming asters. {\bf{(A--C)}} Each row corresponds to a parameter set (A: $w_0=0.15$, $\lambda=1.2$; B: $w_0=0.2$, $\lambda=1.8$; C: $w_0=0.3$, $\lambda=1.6$). Reading from left to right (1) visualization of the data in the two-dimensional simulation domain $[0,128]\times [0,128]$ at the last data point in time (color: number density $\rho$, arrows: polarity density $\bm{W}$); (2) identified clusters of similar dynamics for the $x$-component of $\bm{W}$ (color: cluster index $c_k$); (3) identified clusters of similar dynamics for the $y$-component of $\bm{W}$;
    (4) inferred dominant force-balance models in each cluster ( Filled squares highlight the active components and colors correspond to different clusters);
    {\bf{(D)}} Line profiles of the density $\rho$, the magnitude of the polarity density $|\bm{W}|^2$, the orientational compressibility $\nabla\cdot\bm{W}$, and the density gradient $\nabla\rho$ across the center of the aster in $x$-direction for $w_0=0.15$, $\lambda=1.2$, colored by cluster membership (color bar).}
    \label{fig:results:aster:overview}
\end{figure*}

Next, we consider the self-organized emergence of a defect in the polarity density field $\bm{W}$. These structures, referred to as asters, are also observable in living systems, for example in the mitotic or meiotic spindle \cite{ndlec_self-organization_1997, shelley_dynamics_2016}, and they have also been reconstituted \textit{in vitro} \cite{ndlec_self-organization_1997, surrey_physical_2001} and \textit{in silico} \cite{yan_toward_2022} in minimal systems involving only microtubules and motor proteins with local alignment interactions. The emergence of  topological defects is also increasingly studied, due to their potentially governing contribution to tissue morphogenesis~\cite{hoffmann_theory_2022, wang_patterning_2023}. In the model of Eqs.~\eqref{eq:model:continuoum_model:density} and \eqref{eq:model:continuoum_model:polarity}, the emergence of an aster is characterized by the appearance of a topological defect of index (charge) $-1$ in the polarity density field $\bm{W}$ with a radially symmetric density profile with a maximum at the defect center.
The polarity magnitude $|\bm{W}|^2$ decays exponentially with radial distance to the defect core~\cite{gopinath_dynamical_2012}.

We generate data for three parameterizations of Eqs.~\eqref{eq:model:continuoum_model:density} and \eqref{eq:model:continuoum_model:polarity} that lead to the formation of stable asters. Data is collected until quasi-steady state, as detailed in Appendix \ref{sec:simulation}. Visualizations of the data at the last time point are shown in Fig.~\ref{fig:results:aster:overview}A--C for the three different parameter sets, respectively. Due to the radial symmetry of asters, we find that $K_{\mathrm{aster}}=3$ is sufficient to capture the distinct local regions. Using more clusters breaks the radial symmetry in the zonation, but identifies the same dynamics (see Appendix \ref{sec:exp_details}). The resulting domain decompositions for the $x$ and $y$-components are visualized in the second and third columns of Fig.~\ref{fig:results:aster:overview}A--C. The GMM automatically identifies the two  length scales through which the aster is defined \cite{gopinath_dynamical_2012}: cluster 0 (blue) approximately encloses the core of the aster until $\bm{W}_{\max}$, and cluster 1 (orange) captures the characteristic length of the exponential decay until $|\bm{W}|^2\approx 0$. Finally, cluster 2 (green) contains all points outside the aster in the unordered background. 

The identified locally dominant force balances for the three clusters are shown in the rightmost column of Fig.~\ref{fig:results:aster:overview}A--C with the corresponding Pareto fronts shown in Fig.~\ref{fig:results:aster:pareto}. Similar models are identified across parameterizations. In the aster core, we find a minimal model dominated by the spontaneous polarization terms $a_2$ and $a_4$, as well as the coupling of the flow to the compressibility of the polarity proportional to $\lambda_2$. As such, strong particle interactions in the dense center of the aster lead to the onset of negative compressibility that stabilizes the structure, which can also be seen in the line profiles of Fig.~\ref{fig:results:aster:overview}D. This is analogous to the mechanism found in microscopic models of motor-driven cytoskeletal assemblies, where motor proteins pausing at filament ends generate the active stresses required to pull the filaments together and form asters \cite{yan_toward_2022}. The same has also been observed experimentally \textit{in vitro}, where motor proteins concentrate at the aster center~\cite{surrey_physical_2001, nedelec_dynamic_2001}.

\begin{figure}
    \centering
    \includegraphics[width=\linewidth,trim={0 2.4cm 0 0},clip]{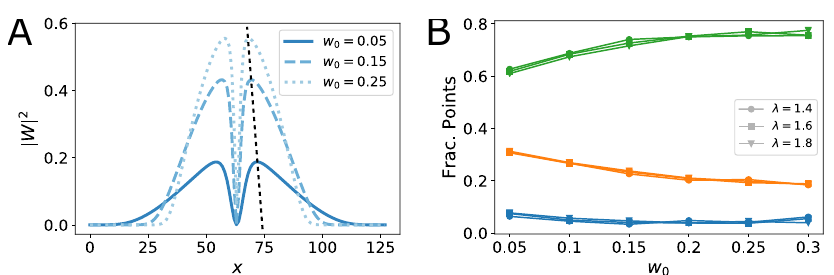}
    \caption{Influence of the self-propulsion speed $w_0$ on the two length scales of the asters at steady state. {\bf{(A)}} Line profile of the squared polarity magnitude $|\bm{W}|^2$ across the center of the asters in $x$-direction for $\lambda=1.6$ and different $w_0$ (inset legend). The black dashed line indicates how the size of the aster core shrinks only minimally for higher $w_0$, while the outer radius of the asters reduces.
    {\bf{(B)}} The fractions of data points assigned to each of the three clusters (colors, see color bar in Fig.~\ref{fig:results:aster:overview}) for three different $\lambda$ (symbols, inset legend) confirms that higher $w_0$ lead to overall smaller asters with similar core sizes, independent of the value of $\lambda$ (symbols, inset legend).}
    \label{fig:results:aster:cluster_ratio}
\end{figure}

In contrast to the moving density bands from Sec. \ref{sec:results:parallel}, however, no sparse support for the model outside of the aster core is found. Therefore, combining all identified terms across clusters recovers the full model, and we find no minimal global model in this case. The domain decomposition given by the identified clusters, however, helps understand the physical mechanisms. In cluster 1, containing the density-decay region of the asters (orange in Fig. \ref{fig:results:aster:overview}), the spontaneous polarization terms $a_2$ and $a_4$ are sufficient to stabilize the ordered state at high densities through strong alignment interactions. The particles are driven toward the high-density region by the self-advection associated to $\lambda_1$ in Eq.~\eqref{eq:model:continuoum_model:polarity}, as well as by the density gradient and the splay-induced pressure proportional to $w_0$ and $\lambda_3$, respectively. This sufficient model closely agrees with previous analytical results \cite{gopinath_dynamical_2012,mishra_fluctuations_2010}, which identified a linear instability where the onset of negative compressibility is controlled by the lowering of the effective pressure of the system due to collective motion for $\lambda_3>0$ \cite{gopinath_dynamical_2012}. The orientational order then causes a flow velocity toward the high-density regions and, thus, a concentration of particles at the aster cores~\cite{marchetti_hydrodynamics_2013}. As the values of all $\lambda_i$ depend on the microscopic interactions of the particles with $\lambda_i\propto w_0^2$ \cite{marchetti_hydrodynamics_2013} it shows how the mechanism directly depends on the particle interactions in the high density region. In the numerical experiments, the self-advection of the particles plays a critical role in the formation of asters. When $\lambda_1=0$, the system undergoes a transition towards the formation of ``streamers'' and vortices \cite{gopinath_dynamical_2012} (see Fig. \ref{fig:results:aster:vortex}), which provides an explanation for the presence of the term in the locally dominant forces of cluster 1. We also observe in numerical experiments (not shown, since in perfect agreement with Gopinath {\it et al.}~\cite{gopinath_dynamical_2012}) that asters cannot form for $w_0=0$, independent of the initialization and the value of $\lambda$. We also confirm that the characteristic length of exponential decay depends on $w_0$, and that this is directly reflected in the identified clusters. In Fig.~\ref{fig:results:aster:cluster_ratio}A, we show that higher $w_0$ shrink the overall size of the aster, but does not significantly influence the size of the core (dashed line). This is also captured by the fraction of data points assigned to ``aster'' clusters (clusters 0 and 1, blue and orange) decreasing proportionally for higher $w_0$ (Fig.~\ref{fig:results:aster:cluster_ratio}B), independent of the value of $\lambda$ (inset legend).

The importance of the local density gradient has also been experimentally observed in aster-forming molecular systems composed of microtubules (MT) interacting with motor proteins. There, the density gradient corresponds to the motor concentration gradient, as motors walk along MT oriented according to $\bm{W}$. Coarse-grained hydrodynamic models of such molecular systems are similar to the model in Eqs.~\eqref{eq:model:continuoum_model:density} and \eqref{eq:model:continuoum_model:polarity}~\cite{sankararaman_self-organized_2004}. It has been demonstrated that in these systems, the stabilization of asters occurs through flows proportional to $\bm{W}$ driven by motors walking along the MT. A motor speed $w_0>0$ thus brings MTs closer together and promotes aster formation~\cite{sankararaman_self-organized_2004}. It has also been pointed out that the steady-state equation for $\bm{W}$ resembles a vector-Poisson equation with a source term proportional to $\nabla\rho$~\cite{gowrishankar_nonequilibrium_2016}. Thus, the non-zero polarity field induces a flow toward the center of the aster that leads to radial alignment of MT. 

Lastly, the dynamics in the dilute unordered background (cluster 2, green in Fig.~\ref{fig:results:aster:overview}) are described by the terms associated to $a_2$, $D_W$,
and $\lambda_3$, where a sub-characteristic density $\rho<\rho_c$ results in $\bm{W}=0$, and $\lambda_3$ plays a similar role as in cluster 1. Interestingly, we observe the appearance of the diffusive relaxation term $\Delta \bm{W}$ in this cluster across all three parameterizations, indicating that the microscopic thermal noise is mostly significant in the unordered phase. In microscopic simulations, the noise has been shown to be important to form larger and denser asters~\cite{yan_toward_2022}. Our results are in agreement with this since diffusion is the dominant transport mechanism for particles in the unordered phase until they get ``drawn in'' by an aster. 

Interestingly, none of the local force-balance laws in the present case contains the time derivative $\partial_t\bm{W}$. We believe that this is the case because the asters form very rapidly, within just a few time points. The dynamics in the data are therefore dominated by the quasi-steady state, where the asters stabilize their size and structure.

\subsection{Density-Dependent Motility}\label{sec:results:density_dependent}

\begin{figure*}
    \centering
    \includegraphics[width=\linewidth,trim={0 10.5cm 0 0},clip]{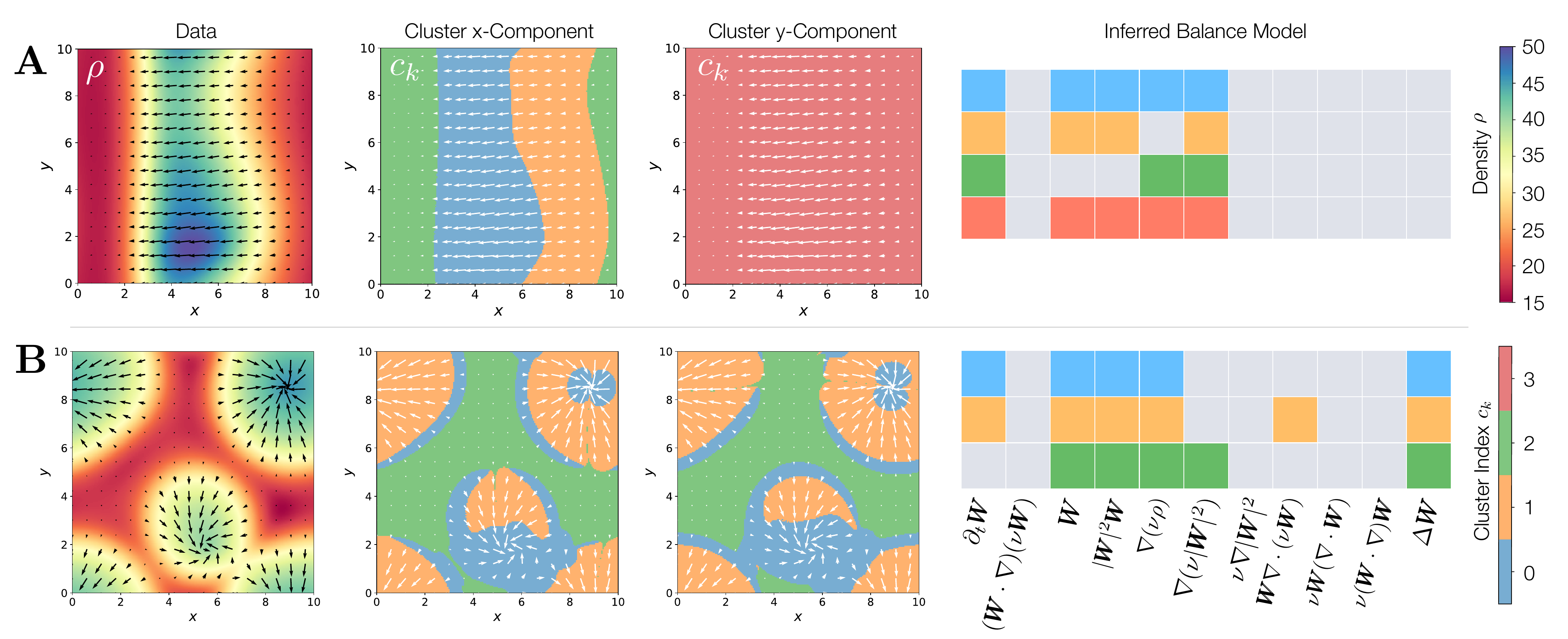}
    \caption{Clusters and the corresponding local dominant models for moving bands and aster with density-dependent motility. {\bf{(A)}} Results for moving density bands, showing from left to right: (1) visualization of the data in the two-dimensional domain $[0,10]\times[0,10]$ (color: number density $\rho$, arrows: polarity density $\bm{W}$) at the last data point in time; (2) identified clusters of similar dynamics for the $x$-component of $\bm{W}$; (3) identified clusters of similar dynamics for the $y$-component of $\bm{W}$; (4) inferred dominant force-balance models in each cluster (Filled squares highlight the active components and colors correspond to different clusters). {\bf{(B)}} Same as A for the case forming an aster.}
    \label{fig:results:density_dependent:overview}
\end{figure*}

In this section, we demonstrate the robustness of our method in identifying universal mechanisms shaping the emergence of complex structures using a more general model that allows for particle motility to be influenced by density \cite{farrell_pattern_2012}. Despite, the slight change in the rules of the microscopic world, the corresponding coarse-grained model exhibits emergent structures like asters and moving bands discussed in the previous section. This provides an opportunity to evaluate if our data-driven approach can reveal common physical principles shaping these structures. For this, it is important that the method is able to efficiently detect potential redundancies in a specified model, regardless of how insignificant their influence might be.

We demonstrate this here by considering a second model of self-propelled particle flows, different from the one presented in Sec.~\ref{sec:method:model}. In this second model, the particle speed depends on the local density as $\nu(n)=v_0\mathrm{e}^{-\lambda n} + v_1$, where $n$ is the number of particles within a given interaction radius. The scalar constants $v_0 \gg v_1 > 0$ define the dilute and crowded limiting velocities. The choice of $\lambda>0$ controls the exponential decay and can lead to the emergence of crowding effects at high densities. It has been shown that the model system is also capable of forming moving density bands and asters~\cite{farrell_pattern_2012}. Here, we consider the system for low $\lambda$ such that the physics are approximately in the same regime and show that the method is capable of identifying the redundant model terms and returns similar physical mechanisms.

Also for this model, a mean-field PDE description has been derived ~\cite{farrell_pattern_2012} for the density field and the local alignment/polarization density field. The dynamics of the mean-field number density $\rho$ are given by:
\begin{equation}\label{eq:results:density_dependent:density}
    \partial_t\rho = -\nabla\cdot(\nu \bm{W} - D_r\nabla\rho) ,
\end{equation}
with diffusion constant $D_r$, while the equation for the polarity density contains additional operators arising from the density-dependence of the speed $\nu(\rho)$:
    \begin{multline} 
        \partial_t \bm{W} + \frac{\gamma}{16\epsilon}(\bm{W}\cdot\nabla)(\nu\bm{W}) = \left(\frac{1}{2}\gamma\rho - \epsilon\right)\bm{W} - 
        \frac{\gamma^2}{8\epsilon}|\bm{W}|^2\bm{W}\\
        -\frac{1}{2}\nabla(\nu\rho) +\frac{3\gamma}{16\epsilon}\nabla(\nu|\bm{W}|^2)- 
        \frac{\gamma}{32\epsilon}\nu\nabla|\bm{W}|^2\\ - \frac{3\gamma}{16\epsilon}\bm{W}\nabla\cdot(\nu\bm{W})-\frac{\gamma}{8\epsilon}\nu\bm{W}(\nabla\cdot\bm{W})\\ - 
        \frac{\gamma}{8\epsilon}\nu(\bm{W}\cdot\nabla)\bm{W}+D_r\Delta\pmb{W}\, ,
        \label{eq:results:density_dependent:polarity}
    \end{multline}
depending on the alignment strength $\gamma$ and the thermal fluctuations $\epsilon$. The diffusion term accounting for the relaxation of the bend and splay moduli is only included for reasons of symmetry and numerical stability. For $\lambda=0$ the additional terms depending on $\nu$ vanish, and the model reduces to the hydrodynamic model in Eq. \eqref{eq:model:continuoum_model:polarity}.

We consider two parameterizations of the model in Eqs.~\eqref{eq:results:density_dependent:density} and \eqref{eq:results:density_dependent:polarity} that form moving density bands and asters visualized in the leftmost panel of Fig. \ref{fig:results:density_dependent:overview}A,B. Compared to the simpler model from the previous sections, the structures are less regular, which is due to the now-present crowding effects. Applying the same data-driven algorithm for inferring locally dominant force balances for the moving density band with $K_{\mathrm{band}}=4$ as before, we find the domain decomposition shown in the remaining panels of Fig.~\ref{fig:results:density_dependent:overview}A. Similar to the case in Sec.~\ref{sec:results:parallel}, cluster 0 (blue) localizes the high-density region of the traveling band and cluster 1 (orange) the flank of the band. The unordered background is also again contained in its own cluster (here cluster 2, green). In contrast to the results in Fig. \ref{fig:results:parallel:overview}, however, we find an additional cluster localized to the $y$-component of the polarity density $\bm{W}_y$. We believe this is due to crowding effects, where the number density in the ordered region starts to form a clump (black arrows) that moves with the band, such that $\partial_t\bm{W}_y\neq 0$. Analyzing the inferred locally dominant balance models in Fig.~\ref{fig:results:density_dependent:overview}C, we find similar mechanics as in Eq. \eqref{eq:results:parallel:minimal_dynamic}, where the high-density region stabilizes and orders due to alignment interactions and gets convected through density gradients (see cluster 2, green) as well as the density-velocity coupling. In contrast to the simpler model considered in the previous sections, however, we do not find a steady-state minimal model, as the data contain no region that can be seen as a steady-state solution in the local reference frame of cluster 0.

\begin{figure}
    \centering
    \includegraphics[width=\linewidth,trim={0 2.9cm 0 0},clip]{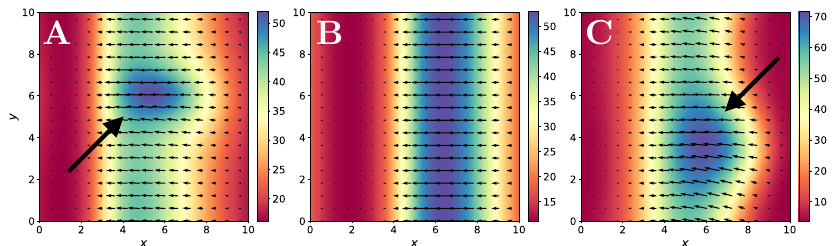}
    \caption{Numerical experiment to validate the origin of crowding clumps in  moving density bands obtained via Eqs. \eqref{eq:results:density_dependent:density} and \eqref{eq:results:density_dependent:polarity}. {\bf{(A)}} Numerical solution of the full model in Eqs. \eqref{eq:results:density_dependent:density} and \eqref{eq:results:density_dependent:polarity}. The crowded density clump is highlighted by the black arrow.
    {\bf{(B)}} Numerical solution of the minimal pattern-forming model in Eq. \eqref{eq:results:parallel:minimal_dynamic} with density-dependent particle speed $\nu(\rho)$,
    forming a moving density band without clump. {\bf{(C)}} Numerical solution of the minimal model as in B, but with the additional term $\nabla(\nu|\bm{W}|^2)$ identified by the data-driven inference (See Fig.~\ref{fig:results:density_dependent:overview}A). This term is sufficient to restore the clumping behavior.}
    \label{fig:results:density_dependent:parallel:validation}
\end{figure}

Interestingly, we also identify $\nabla(\nu|\bm{W}|^2)$ as an additional term driving band motility. This term is similar to the term associated to $\lambda_3$ in Eq. \eqref{eq:model:continuoum_model:polarity} which now also depends on the change in the particle speed $\nu(\rho)$. As such, it not only accounts for the splay deformations but also for the local crowding behavior. We therefore hypothesize that this term governs the crowding effects that lead to the formation of the moving clump in the density band, since this causes gradients in the polarity magnitude $|\bm{W}|^2$. We validate this hypothesis by numerical solutions as shown in Fig. \ref{fig:results:density_dependent:parallel:validation}. Again, the minimal pattern-forming model in Eq.~\eqref{eq:results:parallel:minimal_dynamic} is sufficient to form a moving band structures, even when using a density-dependent particle speed. The band shape is more regular and comparable to that of the model with constant self-propulsion speed. The minimal model can therefore not account for the observed clumping behavior in the full model. Including the term $\nabla(\nu|\bm{W}|^2)$ into the minimal model, however, is sufficient to reproduce the clump crowding behavior observed in the full model. This suggests that initially small gradients in the local particle density $\rho$ will lead to a positive feedback loop, where particles in the denser region slow down. This effect gets amplified through the influx of new particles, which effectively leads to the observed crowding behavior.

We next consider the formation of asters in the model with density-dependent motility. The results are summarized in Fig.~\ref{fig:results:density_dependent:overview}B. We find a similar domain decomposition as in Sec.~\ref{sec:results:parallel} with cluster 0 (blue) capturing the cores of the asters, cluster 1 their shell, and cluster 2 the unordered background. In contrast to Fig.~\ref{fig:results:aster:overview}, the interface region between an aster and the background is also assigned to cluster 0. We believe this is because multiple asters now form in the domain. This is supported by the result in Fig.~\ref{fig:results:density_dependent:aster:top_cluster}, where we only use the right third of the domain to infer the clusters. This portion of the space only contains one aster. Then, cluster 0 is again only assigned to the aster core.

The inferred locally dominant force balances in Fig. \ref{fig:results:density_dependent:overview}D confirm the appearance of $\nabla(\nu\rho)$ and the spontaneous polarization terms, which both were identified as the most important terms in the model with constant particle speed. These same terms have also been found as the key pattern-formation drivers by others, where it was shown that $\nabla(\nu\rho)$ acts as an ordering field for $\bm{W}$ at steady state \cite{farrell_pattern_2012}. It is also responsible for the so-called {\em clustering instability} that leads to the formation of a crowded phase in the asters \cite{farrell_pattern_2012}. In the background cluster 2 we also find $\nabla(\nu|\bm{W}|^2)$, that is similar to $\lambda_3$ in Eq. \eqref{eq:model:continuoum_model:polarity} with additional crowding effects. This suggests that the pattern emerges by a similar mechanism of lowering the effective pressure in the aster cores due to strong particle interactions, which results in the onset of splay-induced negative compressibility. The additional dependence on the gradient of $\nu(\rho)$ indicates, that the process is also influenced by the local crowding behavior. As particles are driven closer together, the speed in the center of the aster approaches $v_1$. This effectively ``traps'' the particles in the region of the core, where the local interactions lead to their alignment. Interestingly, we observe the phenomenological appearance of the coupling to the compressibility of the polarity only in the shell regions, but not in the cores, of the asters. This, however, might be an artifact of the outer parts of the shell being assigned to cluster 0, since in the cropped case with a single aster (see Fig. \ref{fig:results:density_dependent:aster:top_cluster}), $\bm{W}\nabla\cdot(\nu\bm{W})$ appears directly after the spontaneous polarization terms controlling the order-disorder transition.

Taken together, these observations show that despite the mathematical differences between the models in Eqs.~\eqref{eq:model:continuoum_model:density} and \eqref{eq:model:continuoum_model:polarity} and in Eqs. \eqref{eq:results:density_dependent:density} and \eqref{eq:results:density_dependent:polarity}, the emerging spatio-temporal structures form by similar mechanisms and how the method correctly ignored potentially redundant terms.
In both models, propagating bands are formed by local alignment interactions driven by gradients in number density, while steady-state asters are shaped by a mechanism of splay-induced negative compressibility arising from strong particle interactions.

\section{Discussion and Outlook}\label{sec:discussion}

We extended the method of Callaham {\it et al.}~\cite{callaham_learning_2021} by higher-dimensional model selection, leading to Pareto fronts with a clear selection point. We demonstrate that the resulting algorithm is able to infer locally dominant components of a global model describing active particle systems. Using a combination of unsupervised clustering with Gaussian Mixture Models (GMM) and sparse principle components analysis (SPCA), we decompose space and time domains into regions exhibiting similar dynamics. 

We looked at the dynamics of two distinct non-equilibrium structures that can also be observed in \textit{in-vitro} experimental studies, namely moving density bands and asters. Our findings indicate that moving density bands are sustained through the dynamic interplay between spontaneous polarization components leading to local particle alignment, and density gradients that stably propel the band. On the other hand, asters are shaped through negative compressibility generated by strong particle alignment interactions. The resulting orientational order on the macroscale then leads to flow towards high-density regions, drawing in more particles.  Furthermore, comparing the identified mechanisms with those obtained for a more detailed model of an active particle system with density-dependent motility enabled us to identify physical commonalities across the two models and shows that the method effectively ignores the redundant terms.

Importantly, the results obtained here only required data from numerical solutions of known hydrodynamic mean-field models. Yet, decomposing them into regions of similar local dynamics and inferring minimal models of the locally dominant forces, we were able to bridge the gap back to the microscopic origins of the involved processes and provide mechanistic explanations for several previous observations in the literature, both theoretical and experimental. We therefore believe that the presented method ideally complements classic asymptotic or linear stability analysis to reason about local dominant force balances, pattern-forming instabilities, and underlying physical mechanisms in self-organized active matter.

Naturally, data-driven approaches are limited by the identifiability of the model given the data. For example, we found that for very small microscopic self-propulsion velocities, the influence of the density gradient $\nabla\rho$ asymptotically vanishes. In those cases, we find exponentially decaying Pareto fronts for the clusters, rendering model selection challenging. Overall, we observed a correlation between the robustness of the data-driven inference and the magnitude of splay deformations. The method is more robust in regions of parameter space that contain pronounced dynamic instabilities. It is therefore important in practice to always check the Pareto fronts of the inference algorithm before interpreting the results. In the models discussed in this work, the hydrodynamic interactions between the particles are ignored. Introducing such interactions is a promising direction to explore, given that many microswimmers generate vorticity in their vicinity in order to move, thereby affecting neighboring particles \cite{marchetti_hydrodynamics_2013}.

Nothwithstanding these limitations of the present approach, we are convinced that data-driven modeling approaches hold great potential for enhancing our mechanistic understanding of active matter under non-dry conditions. Especially the combination with classic first-principles modeling and other data-driven methods~\cite{maddu_stability_2022,Maddu:2021a} holds great potential for getting closer to a multi-scale understanding of the physics of morphogenesis and pattern formation in living biological systems directly from experimental data.

\section*{Data Availability}
The data, models, and source codes that support the findings of this study are openly available at the following URL/DOI: \url{http://doi.org/10.14278/rodare.2378}

\section*{Acknowledgments}
This work was partly funded by the Center for Advanced Systems Understanding (CASUS), which is financed by Germany's Federal Ministry of Education and Research (BMBF) and by the Saxon Ministry for Science, Culture and Tourism (SMWK) with tax funds on the basis of the budget approved by the Saxon State Parliament. This work was supported by the German Research Foundation (Deutsche Forschungsgemeinschaft, DFG) under Germany’s Excellence Strategy -- EXC-2068-390729961 -- Cluster of Excellence “Physics of Life” of TU Dresden, and by the Center for Scalable Data Analytics and Artificial Intelligence (ScaDS.AI) Dresden/Leipzig, funded by the Federal Ministry of Education and Research (Bundesministerium f\"{u}r Bildung und Forschung, BMBF).

\appendix

\section{Numerical Solver Details}\label{sec:simulation}

We present the details of the numerical algorithms and their parameterizations used to generate the data from the numerical solution of the two models studied here.

\subsection{Hydrodynamic Model of Self-Propelled Particles}

The results presented in Sec.~\ref{sec:results:parallel} and \ref{sec:results:aster} use data generated by numerically solving the hydrodynamic model in Eqs. \eqref{eq:model:continuoum_model:density} and \eqref{eq:model:continuoum_model:polarity}. We use the same parameter configurations as Gopinath {\it et al.}~\cite{gopinath_dynamical_2012}, which specifically includes setting $\lambda_1=\lambda_2=\lambda_3=:\lambda$ as well as $D=D_W=1$. This allows us to use the published phase diagrams for this case~\cite{gopinath_dynamical_2012} as guidance in determining the parameter sets for which bands and asters are expected to form. For the spontaneous polarization terms, we use the form $a_2(\rho)=1-\rho/\rho_c$ and $a_4(\rho)=1+\rho/\rho_c$ with $\rho_c=1$, following Mishra {\it et al.}~\cite{mishra_fluctuations_2010}.

The governing equations are numerically solved in the OpenFPM framework for scalable scientific computing~\cite{incardona_openfpm_2019} using fourth-order finite differences on a two-dimensional regular Cartesian grid of $128\times 128$ points for the asters and $256\times 92$ points for the moving density bands with resolution $\Delta x=1$ and periodic boundary conditions in all directions. Time integration is done using the explicit Euler method with an additional correction step based on the trapezoidal rule with time-step size $\Delta t=0.02$. Data is generated and stored every 2500 simulation time steps. The time $\Delta t_{\text{data}}=2500\Delta t$ is the {\em data time step} used for the data-driven inference. The initial conditions are: $\rho_0\mathcal{U}\left(-\pi, \pi\right)$ with initial density $\rho_0=1.05$ and $\mathcal{U}$ a uniform distribution, and $\bm{W}(t=0)=\bm{0}$.

Further, we use the following parameters:
\paragraph{Density Bands:} The data consists of 21 data time points in the dynamic regime of the moving density band with $T_{\text{start}}=68\Delta t_{\text{data}}$ and $T_{\text{end}}=88\Delta t_{\text{data}}$. The model parameters are: 1) $w_0=0.5$, $\lambda=0.2$; 2) $w_0=0.4$, $\lambda=0.4$; 3) $w_0=0.25$, $\lambda=0.25$.
\paragraph{Asters:} We collect data time points until the systems reaches a quasi-steady state, defined by $\partial_t\bm{W}\leq 10^{-4}$, or a maximum of 200,000 simulation time steps. The model parameters are:
1) $w_0=0.15$, $\lambda=1.2$; 2) $w_0=0.2$, $\lambda=1.8$; 3) $w_0=0.3$, $\lambda=1.6$. 

We construct the feature matrix in Eq.~\eqref{eq:appendix:methods:feature_matrix} from the numerical solution data by approximating all spatial derivatives  using forth-order finite differences. The first column, $\partial_t\bm{W}$, is obtained directly from the corrector step of the numerical solver.

\subsection{Model with Density-Dependent Motility}

The results presented in Sec.~\ref{sec:results:density_dependent} use data generated by numerically solving Eqs. \eqref{eq:results:density_dependent:density} and \eqref{eq:results:density_dependent:polarity} for the parameters studied by Farrell {\it et al.}~\cite{farrell_pattern_2012}. The equations are numerically solved in the OpenFPM framework \cite{incardona_openfpm_2019} using second-order finite differences over the two-dimensional domain $\Omega=[0,10]^2$ discretized with a regular Cartesian grid of $200\times 200$ points. Periodic boundary conditions are used in both directions. Time integration uses an explicit Euler scheme with a corrector step based on the trapezoidal rule with time step $\Delta t=10^{-3}$. Data for inference was stored every 500 time steps, thus $\Delta t_{\text{data}}=500\Delta t$. The  density at time 0 is constant $\rho_0=30$, and the initial condition for the polar density is: $\bm{W}(t=0)=\rho_0(\cos\zeta,\, \sin\zeta)^\top$, $\zeta\sim\mathcal{U}(-\pi,\pi)$. We compute the speed as $\nu(\rho)=v_0\exp\{-\lambda\pi R_0^2\rho\} + v_1$. The feature matrix $\Theta$ is constructed as before, but with second-order finite differences.

Further, we use the following parameters:
\paragraph{Density Bands:} The data consists of 21 data time points after the density band has formed with $T_{\text{start}}=89\cdot\Delta t_{\text{data}}$ and $T_{\text{end}}=109\cdot\Delta t_{\text{data}}$. The model parameters are: $v_0 = 2$, $v_1 = 0.1$, $\lambda = 10^{-5}$, $\gamma = 0.16$, $R_0 = 1$, $\epsilon = 2$, $D_r = 0.4$.
\paragraph{Asters:} The data consists of 21 data time points in the quasi-steady state between $T_{\text{start}}=370\cdot\Delta t_{\text{data}}$ and $T_{\text{end}}=390\cdot\Delta t_{\text{data}}$. The model parameters are: $v_0 = 2$, $v_1 = 0.0$, $\lambda = 0.05$, $\gamma = 0.16$, $R_0 = 1$, $\epsilon = 2.08$, $D_r = 0.02$.

\section{Experimental Details}\label{sec:exp_details}

\begin{figure}
    \centering
    \includegraphics[width=.8\linewidth,trim={0 3.2cm 0 0},clip]{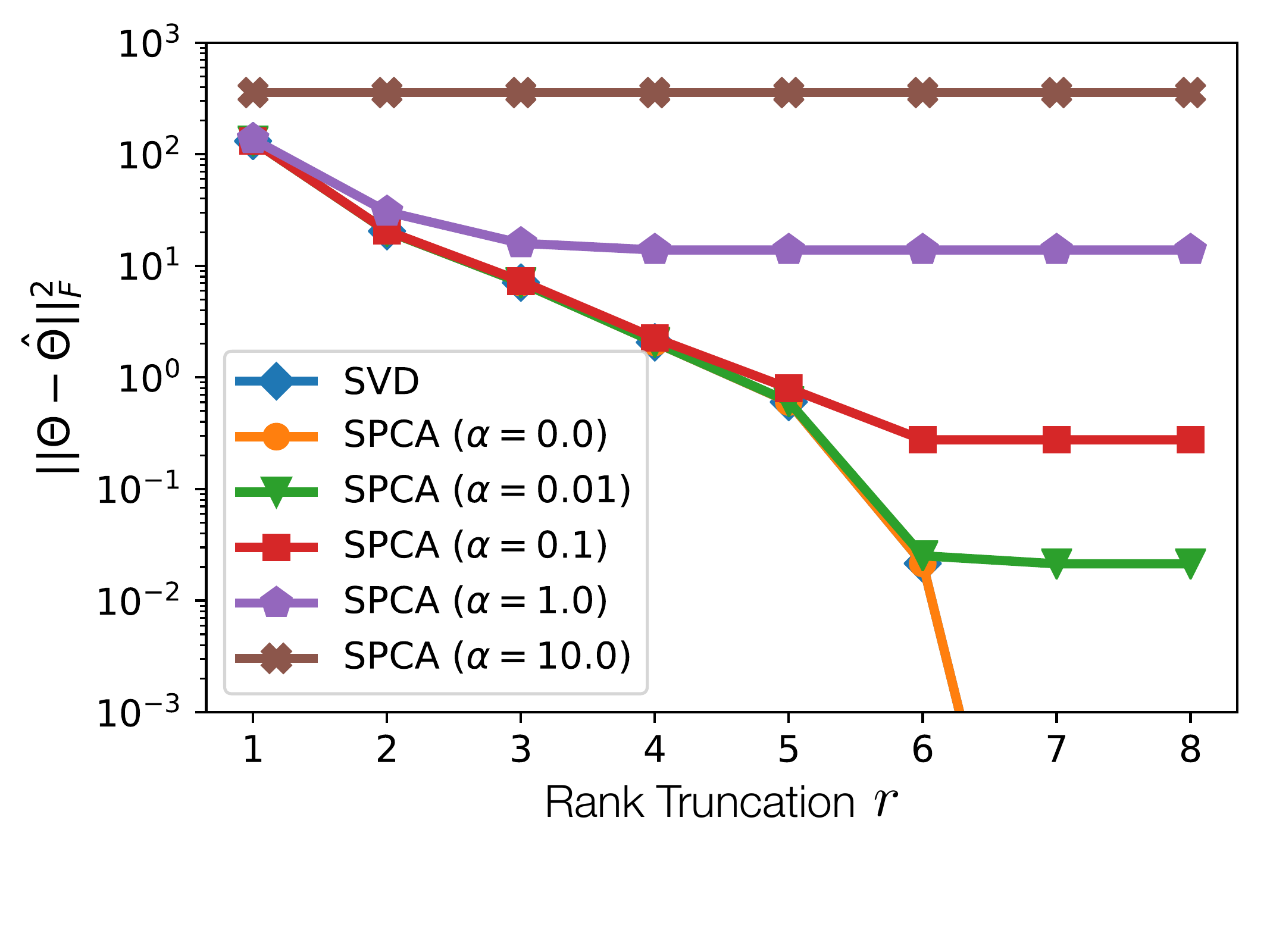}
    \caption{Reconstruction error for different rank-$r$ truncations by minimizing the objective in Eq.~\eqref{eq:appendix:methods:objective} over the whole domain for $w_0=0.05$ and $\lambda=1.6$ (aster case). Different lines correspond to different choices for the sparsity-promoting parameter $\alpha$ in SPCA (inset legend). Blue diamonds give the reference reconstruction error of rank-$r$ truncated SVD, which perfectly coincides with SPCA for $\alpha=0$ and reconstructs to machine precision with full-rank approximation (not shown), validating the correctness of the SPCA algorithm implementation.}
    \label{fig:exp_details:reconstruction}
\end{figure}

\begin{figure}
    \centering
    \includegraphics[width=.9\linewidth]{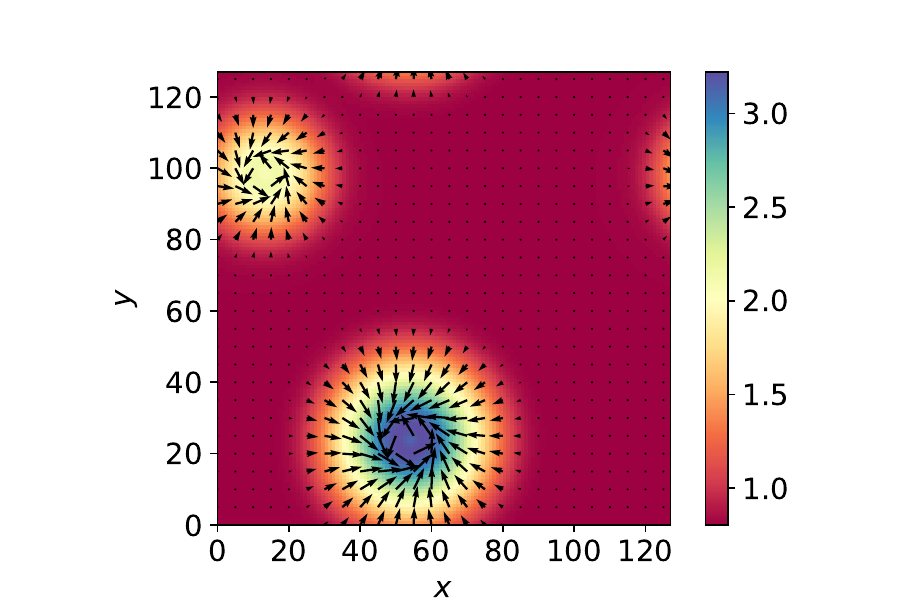}
    \caption{Numerical solution of the hydrodynamic equations in Eqs. \eqref{eq:model:continuoum_model:density} and \eqref{eq:model:continuoum_model:polarity} with $w_0=0.2$ and $\lambda=1.8$ (same as Fig. \ref{fig:results:aster:overview}B) but setting $\lambda_1=0$. The quasi-steady state solution (see Appendix \ref{sec:simulation}) forms a vortex instead of an aster.}
    \label{fig:results:aster:vortex}
\end{figure}

\begin{figure}[t]
    \centering
    \includegraphics[width=\linewidth]{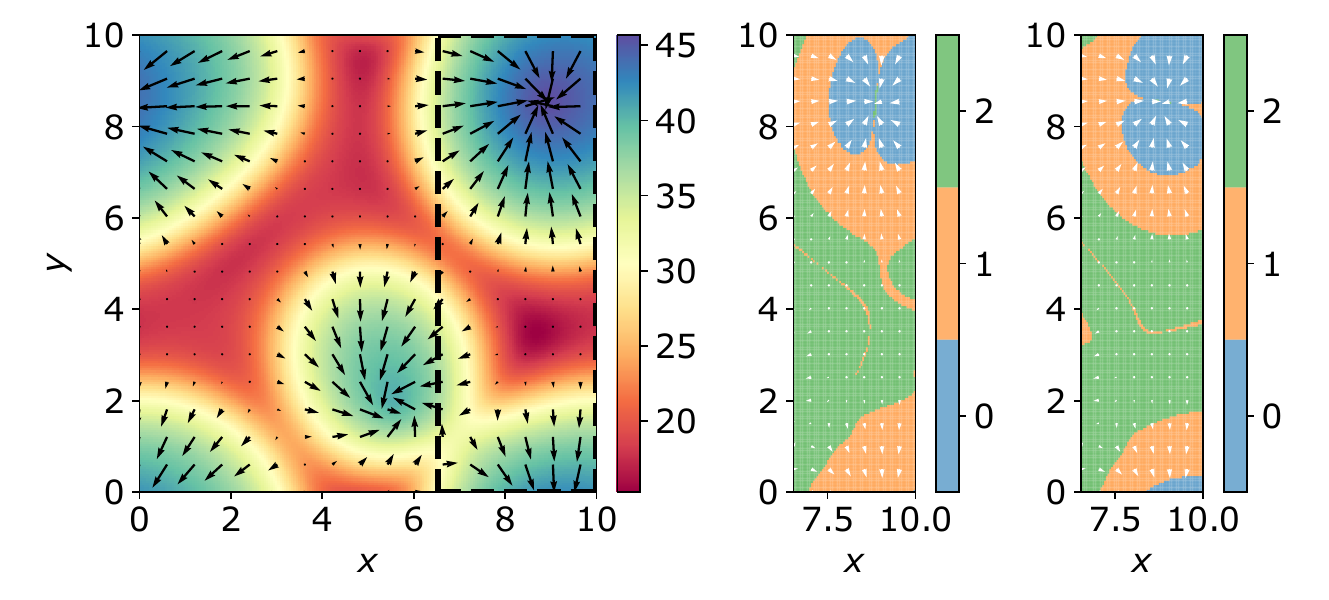}
    \caption{Inferred clusters for a single aster in the density-dependent motility model. The data is cropped to contain only the right third of the domain with a single aster. The left panel shows the original data with the black dashed box indicating the region that is used. The center and right panels show the inferred clusters for the $x$ and $y$-components of $\bm{W}$, respectively. 
    Decreasing $\alpha$ in the SPCA, we observe appearance of terms in the following order: $(\frac{1}{2}\gamma\rho-\epsilon)\bm{W}$, $\frac{\gamma^2}{8\epsilon}|\bm{W}|^2\bm{W}$, $\Delta\bm{W}$, and $\frac{3\gamma}{16\epsilon}\bm{W}\nabla\cdot(\nu\bm{W})$ (corresponding to $\lambda_2$).}
    \label{fig:results:density_dependent:aster:top_cluster}
\end{figure}

We provide the hyperparameters of all components of the machine-learning framework used in this paper. We always choose 1 as the random seed for both the GMM and SPCA. All computer codes used are freely available at the following URL/DOI: \url{http://doi.org/10.14278/rodare.2378}.

\paragraph{Gaussian Mixture Model:} The GMM is trained with three reinitializations until convergence with tolerance $10^{-3}$ and a maximum of 100 iterations (which in our experiments was never reached) using the expectation-maximization (EM) algorithm \cite{cherkassky_learning_2007} implemented in $\textsc{sklearn}$ \cite{pedregosa_scikit_2011}. The centers of the clusters are initialized by $k$-means clustering. The number of clusters $K$ is chosen based on the desired spatial decomposition as $K_{\mathrm{aster}}=3$ and $K_{\mathrm{band}}=4$ for both hydrodynamic models. Higher numbers result in finer domain decompositions with the same spatial characteristics. As such, they do not yield additional information about the physics of the system. Algorithmically, $K$ could therefore be determined in a post-processing step, where initially too-many clusters are merged if the identified dominant force balances within them are equal \cite{callaham_learning_2021}. In the present work, however, we fixed the number of clusters manually from prior knowledge.

\paragraph{Sparse PCA:} We compute the first $r=6$ principle components for each cluster $\Theta_k$, $k=0,\ldots , K-1$. The rank-$r$ truncation is chosen based on the reconstruction error. This is illustrated in Fig.~\ref{fig:exp_details:reconstruction}, where the reconstruction error (as the square of the Frobenius norm) is plotted versus $r$ for different values of the sparsity-promoting SPCA parameter $\alpha$. For small $\alpha>0$, we find an elbow in the curve at $r=6$, which we use for the experiments throughout the paper. Higher regularizations $\alpha$ move the elbow to lower $r$, limiting the possible reconstruction error. We use $r=6$ to ensure correct asymptotic behavior over a wide range of regularizations during model selection.
We also find that the choice of $r$ does not significantly influence the model paths or the identified minimal models. Higher $r$ mainly improve the possible reconstruction error (limited by the SVD), leading to clearer selection points in the Pareto front.
We minimize the objective in Eq.~\eqref{eq:appendix:methods:objective} using alternating coordinate descent, for which we observe lower final reconstruction errors and better-resolved Pareto fronts than for the least-angle regression (LARS) optimizer~\cite{efron_least_2004}, albeit at higher computational cost. We use the implementation from $\textsc{sklearn}$ \cite{pedregosa_scikit_2011} with an absolute tolerance of $10^{-6}$ and a maximum of 1000 iterations (which was never reached in our experiments).

\begin{figure*}
    \centering
    \includegraphics[width=\linewidth]{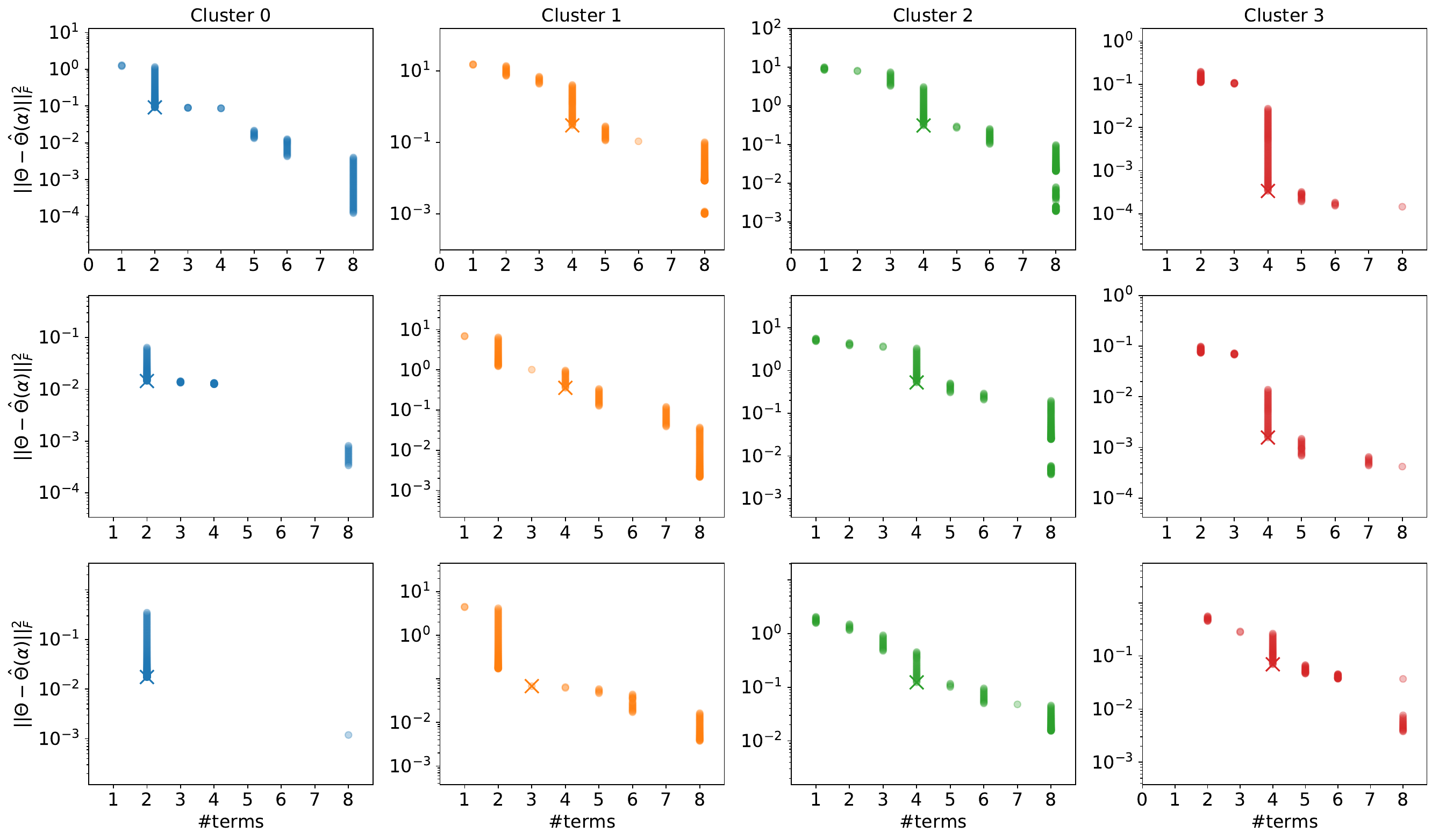}
    \caption{Pareto fronts for moving density bands used for model selection in Sec. \ref{sec:results:parallel}. The selected points are indicated by a cross. The columns from left to right correspond to the different spatial clusters as indicated in the column headings. The rows from top to bottom correspond to the three parameterizations: 1) $w_0=0.5$, $\lambda=0.2$; 2) $w_0=0.4$, $\lambda=0.4$; 3) $w_0=0.25$, $\lambda=0.25$.}
    \label{fig:results:parallel:pareto}
\end{figure*}

\begin{figure*}
    \centering
    \includegraphics[width=.75\linewidth]{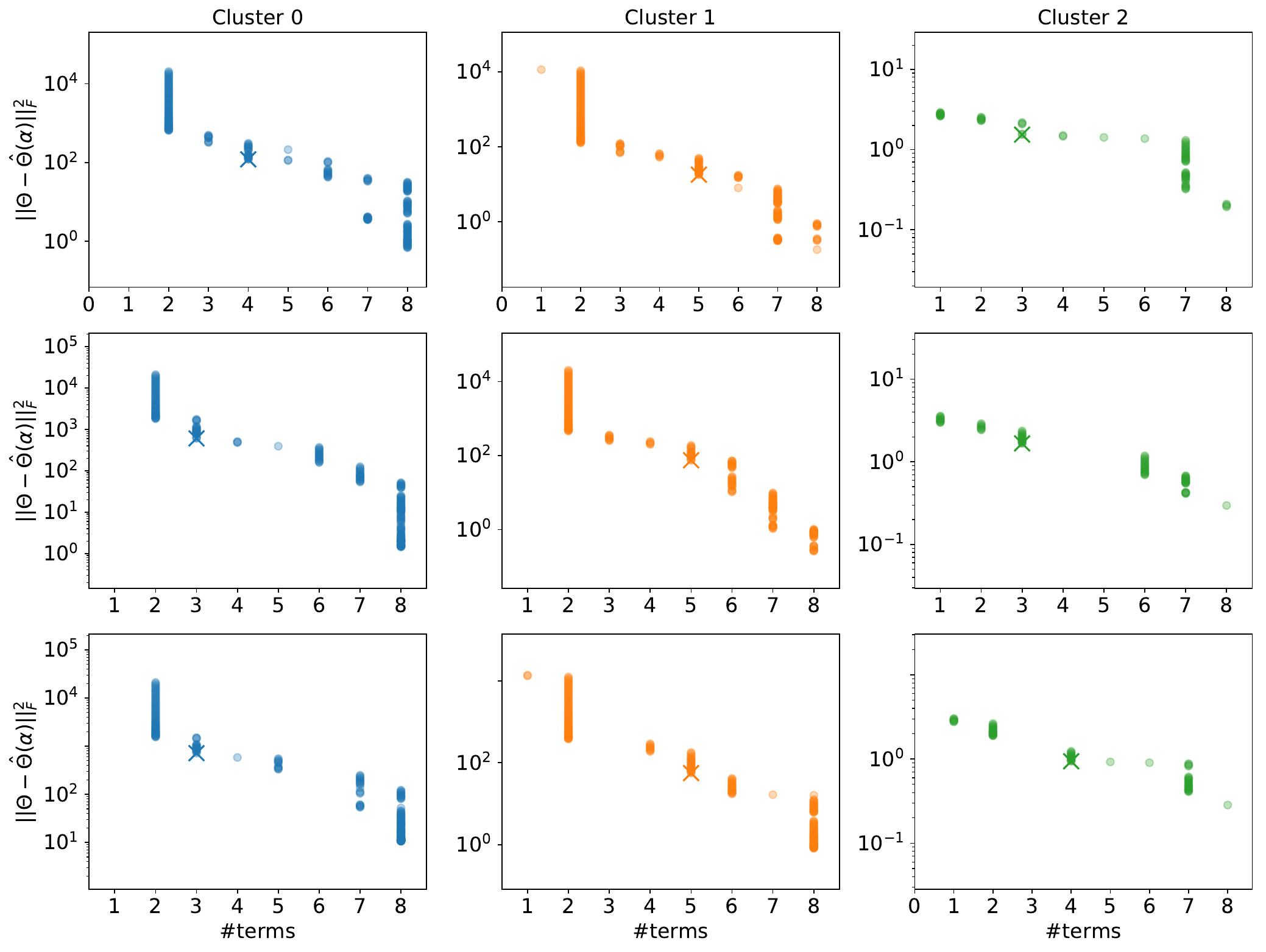}
    \caption{Pareto fronts for asters used for model selection in Sec. \ref{sec:results:aster}. The selected points are indicated by a cross. The columns from left to right correspond to the different spatial clusters as indicated in the column headings. The rows from top to bottom correspond to the three parameterizations: 1) $w_0=0.15$, $\lambda=1.2$; 2) $w_0=0.2$, $\lambda=1.8$; 3) $w_0=0.3$, $\lambda=1.6$. The increased reconstruction errors are due to the higher absolute values on the data points.}
    \label{fig:results:aster:pareto}
\end{figure*}

\begin{figure*}
    \centering
    \includegraphics[width=\linewidth,trim={0 12cm 0 0},clip]{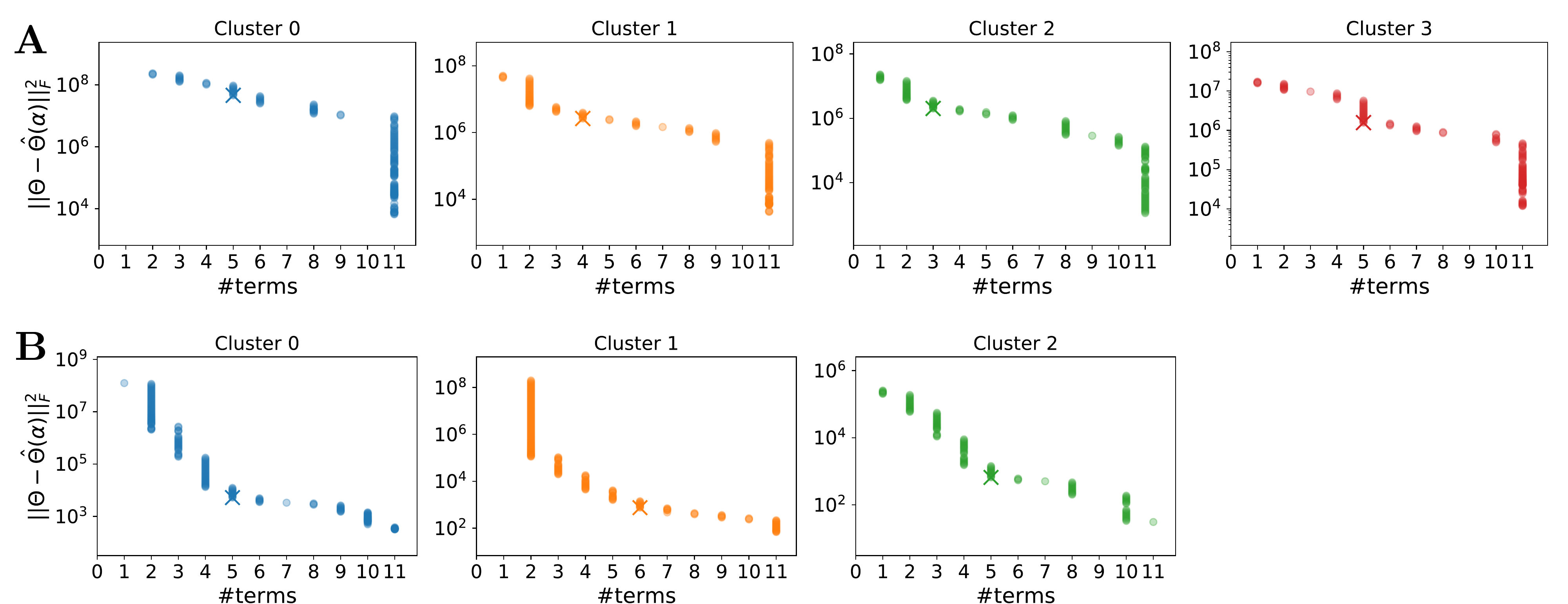}
    \caption{Pareto fronts for the model with density-dependent motility used for model selection in Sec. \ref{sec:results:density_dependent}. The selected points are indicated by a cross. The columns from left to right correspond to the different spatial clusters as indicated in the column headings. {\bf{(A)}} Pareto fronts for moving density bands. {\bf{(B)}} Pareto fronts for asters. The increased reconstruction errors are due to the higher absolute values on the data points.}
    \label{fig:results:density_dependent:pareto}
\end{figure*}

\paragraph{Pareto Fronts:} Pareto fronts are obtained for inspection by evaluating the residual of Eq. \eqref{eq:appendix:methods:objective} on a log-equidistant grid of 400 points over $\alpha\in[10^{-4},10^2]$. The only exception is Sec. \ref{sec:results:density_dependent}, where we use $\alpha\in[10^{-4},10^4]$ to obtain the full Pareto front. The residual is plotted versus the number of terms in $\hat{M}_k(\alpha)$. We then select the point on the Pareto front that has the steepest descent with at least two two non-zero components, since a force balance requires at least two terms. If there is no such model, we use ``elbow'' selection. The Pareto fronts used for model selection in Sec. \ref{sec:results:parallel} are shown in Fig. \ref{fig:results:parallel:pareto}, those for Sec. \ref{sec:results:aster} in Fig. \ref{fig:results:aster:pareto}, and those for Sec. \ref{sec:results:density_dependent} in Fig. \ref{fig:results:density_dependent:pareto}.

%\bibliography{apssamp}


\end{document}